\documentclass{aa1}
\usepackage{graphicx}
\usepackage[varg]{txfonts}
\usepackage{cases}
\usepackage{natbib}

\begin{document}

\title{Double-periodic pulsations simultaneously detected in mid-infrared and hard X-ray emissions during an X1.5 flare}

\author{Dong~Li \inst{1}, Yuyang~Ye \inst{2,3}, Xianyong~Bai \inst{2}, Xu~Yang \inst{4}}

\institute{Purple Mountain Observatory, Chinese Academy of Sciences, Nanjing 210023, PR China \email{lidong@pmo.ac.cn} \\
           \and State Key Laboratory of Solar Activity and Space Weather, National Astronomical Observatories, Chinese Academy of Sciences, Beijing 100101, PR China \email{xybai@bao.ac.cn} \\
           \and School of Astronomy and Space Science, University of Chinese Academy of Sciences, Beijing 100049, PR China \\
           \and Xinjiang Astronomical Observatory, Chinese Academy of Sciences, Urumqi 830011, PR China\\}

\date{Received; accepted}

\titlerunning{Double-periodic pulsations in solar mid-infrared emissions}
\authorrunning{Dong Li et al.}

\abstract {Quasi-periodic pulsations (QPPs) have been observed in a
broad electromagnetic spectrum, encompassing radio, ultraviolet,
white light, X-rays, and $\gamma$-rays. Yet, flare-associated QPPs
remain rarely detected in mid-infrared emission.} {We explored
dual-period QPPs in the mid-infrared waveband, hard X-rays (HXRs),
and microwave emission during an X1.5 flare on 2024 December 30
(SOL2024-12-30T04:01).} {The X1.5 flare was captured simultaneously
by the Accurate Infrared Magnetic field Measurements of the Sun
(AIMS), Fermi, the Hard X-ray Imager (HXI), the Nobeyama Radio
Polarimeters (NoRP), the White-light Solar Telescope (WST), and the
Atmospheric Imaging Assembly (AIA). Dual quasi-periods were
identified using fast Fourier transform and wavelet analysis
methods. Hard X-ray images were reconstructed with the HXI\_CLEAN
algorithm. Differential emission measure (DEM) describes the
temperature structure of optically thin plasmas in the solar corona,
and its solution was applied in six AIA Extreme ultraviolet
wavelengths.} {Flare QPPs with dual quasi-periods at $\sim$8.5~s and
$\sim$4.6~s were simultaneously detected in the AIMS~8-10~$\mu$m,
Fermi~26$-$50~keV, and HXI~20$-$50~keV wavebands during the
impulsive phase. Imaging observations show that the flare emission
sources in the mid-infrared, HXR, and white-light wavebands are
spatially coincident. Mid-infrared emissions are primarily localized
at the loop top and double footpoints, which are connected by a hot
plasma loop. These observational features support intermittent and
rapid energy release via oscillatory magnetic reconnection during
the solar flare. Differential emission measure analysis confirms
fast sausage waves in flaring loops, while the WST observation
indicates a white-light flare.} {We localized the flare QPPs with
double periods in mid-infrared, HXR, and microwave emissions during
a white-light flare. The flare-associated QPPs may be attributed to
a quasi-periodic regime of magnetic reconnection, with the double
periods likely modulated by quasi-harmonics of sausage waves.}


\maketitle

\section{Introduction}
Solar flares are among the most energetic events on the solar disk,
releasing energy up to 10$^{32}$~ergs within a few minutes.
Consequently, they often exhibit violent, rapid and impulsive
characteristics \citep{Priest02,Fletcher11}. The eruption of solar
flares can occur across almost all electromagnetic radiation bands,
and they are readily observable in wavelengths spanning
radio/microwave, ultraviolet (UV), white light, soft/hard X-rays
(SXR/HXR) and even $\gamma$-rays \citep{Benz17}. However, infrared
emissions from solar flares remain infrequently documented, and
their properties are poorly understood, despite the potential for
detection in this range to provide a unique window into the upper
photosphere and lower chromosphere \citep{Ohki75,Kaufmann13,Yang25}.

Quasi-periodic pulsations (QPPs) represent a common phenomenon,
featuring strongly variable modulations in flare emissions. A
typical QPP event is usually characterized by multiple successive
and impulsive pulses in time-dependent intensity curves of solar
flares \citep[e.g.,][]{McLaughlin18,Kupriyanova20,Zimovets21}.
Quasi-periodic pulsations often display diverse temporal
characteristics and plasma emission signatures shaped by conditions
in the flare region. As such, they serve as an important diagnostic
tool for probing coronal parameters and energy-release processes in
solar active regions. They may be associated with oscillatory
magnetic reconnection, repeated particle acceleration, and
magnetohydrodynamic (MHD) waves
\citep[e.g.,][]{Li24a,Li25a,Purkhar25,Shabalin25}. Therefore, the
study of QPPs continues to be a subject of ongoing interest and has
grown in importance \citep{Tan08,Reale26}. To date, QPPs in solar
flares have been observed across nearly the entire electromagnetic
spectrum, i.e., radio/microwave, white light, H$\alpha$, Ly$\alpha$,
UV, Extreme ultraviolet (EUV), SXR/HXR, and even $\gamma$-rays
\citep[e.g.,][]{Nakariakov10,Nakariakov18,Shen22,Li23,Li24b,Li25b,Li26a,Collier24,Lim25,Song25,Cattell26}.
However, flare QPPs are rarely observed in infrared emission,
largely due to insufficient observational data in this spectral
range.

A classical QPP event typically exhibits more than three successive
pulses in the flare intensity curve, though this criterion may not
serve as a strict definition \citep{McLaughlin18,Inglis24}. The term
`quasi-period' denotes the average duration of pulses within a
single QPP event, as observed pulse timescales are often unequal or
irregular \citep{Kupriyanova20}. The quasi-period of flare QPPs
spans a broad range of timescales, from a few seconds to tens of
minutes
\citep[e.g.,][]{Tan10,Tan16,Kolotkov18,Carley19,Li22,French25,Shi25,Szaforz25,Zhang26}.
Quasi-periodic pulsations with multiple quasi-periods have been detected within
individual flares \citep[e.g.,][]{Li21,Karlicky23,Song25},
particularly the double-periodic cases \citep{Li17,Ashfield25}.
While the ratio between dual periods frequently differs
significantly from 2, this deviation may arise from highly
dispersive MHD modes \citep{Nakariakov05} or other physical effects
such as longitudinal density stratification \citep{Andries05}. In
the literature, flare QPPs are typically associated with MHD waves,
such as slow magnetoacoustic waves \citep{Wang21}, fast sausage
waves \citep{Li20}, and kink waves \citep{Nakariakov21}. They may
also be triggered by periodic regimes of magnetic reconnection
\citep{Kumar24,Zhu25,Zimovets25,Schiavo26}. Magnetohydrodynamic waves can directly
drive fluctuations in local densities and temperatures within flare
cores \citep{Yuan15}, while oscillatory magnetic reconnection may
periodically accelerate nonthermal electrons, resulting in
quasi-periodic precipitations of high-energy particles
\citep{Ashfield25,Li26}. Even though some of the physical mechanisms
capable of producing QPPs are associated with particular
observational characteristics (e.g., typical period ranges),
identifying the generation mechanism in a specific QPP event often
remains an open question, largely because of insufficient
observational data \citep[cf.][]{Kupriyanova20}.

The Accurate Infrared Magnetic field Measurements of the Sun
\citep[AIMS;][]{Deng25}, a newly built ground-based telescope,
enables the investigation of flare QPPs in mid-infrared emissions and
facilitates pinpointing their origin. This article focuses on
double-periodic QPPs observed simultaneously in mid-infrared and HXR
emissions during an X1.5 flare.

\section{Observations}
We analysed an X1.5 flare that occurred on 2024 December 30 in the
active region of NOAA~13936. The event was simultaneously measured
by AIMS, Fermi \citep{Meegan09}, the Hard X-ray Imager
\citep[HXI;][]{Su19}, the White-light Solar Telescope
\citep[WST;][]{Feng19} of the Advanced Space-based Solar Observatory
\citep[ASO-S;][]{Gan19,Huang19}, the Atmospheric Imaging Assembly
\citep[AIA;][]{Lemen12}, the Helioseismic and Magnetic Imager
\citep[HMI;][]{Schou12} Instrument aboard the Solar Dynamics
Observatory (SDO), the Nobeyama Radio Polarimeters (NoRP), the X-ray
Sensor (XRS) on board Geostationary Operational Environmental
Satellite-16 (GOES-16), and the Spectrometer/Telescope for Imaging
X-rays \citep[STIX;][]{Krucker20} on board Solar Orbiter (SolO). At
the time of the X1.5 flare, SolO was located at 13.7$^{\circ}$ east
in solar longitude from the Earth-Sun line, and it was very close to
Earth (at approximately 0.95~AU from the Sun), corresponding to a
timing difference of about 14.6~s relative to Earth-based
observations. The level~1 datasets were downloaded from the homepage
of each instrument for subsequent analysis. A comprehensive
summary of the employed instruments and their key parameters is
presented in Table~\ref{tab1}.

\begin{table*}[ht]
\centering \caption{Instruments and parameters used in this work.}
\label{tab1}
\begin{tabular}{cccccc}
\hline
Instrument &  Waveband      &  Cadence   &  Description   &  Pixel scale          &  FOV        \\
\hline
AIMS       &  8-10~$\mu$m   &   1~s      &  mid-infrared  & 1.5$^{\prime\prime}$  & 384$^{\prime\prime}$$\times$384$^{\prime\prime}$        \\
Fermi      &  26-50~keV     &   1.024~s  &  HXR           &    ---                &  Full-disk   \\
ASO-S/HXI  &  10-50~keV     &   1~s      &  SXR           &  2$^{\prime\prime}$   &  Full-disk   \\
ASO-S/HXI  &  20-50~keV     &   1~s      &  HXR           &  2$^{\prime\prime}$   &  Full-disk   \\
ASO-S/WST  &  3600~{\AA}    &  120~s     &  WL            &  0.6$^{\prime\prime}$ &  Full-disk   \\
SolO/STIX  &  25-50~keV     &   3~s      &  HXR           &    ---                &  Full-disk   \\
GOES-16/XRS &  1-8~{\AA}    &   60~s     &  SXR           &    ---                &  Full-disk   \\
NoRP       &    2~GHz       &   1~s      &  Microwave     &    ---                &  Full-disk   \\
SDO/AIA    &   131~{\AA}    &   12~s     &  EUV           & 0.6$^{\prime\prime}$  &  Full-disk   \\
SDO/AIA    &  1600~{\AA}    &   24~s     &  UV            & 0.6$^{\prime\prime}$  &  Full-disk   \\
SDO/AIA    &  1700~{\AA}    &   24~s     &  UV            & 0.6$^{\prime\prime}$  &  Full-disk   \\
SDO/HMI    &  M             &   45~s     &  LOS           & 0.5$^{\prime\prime}$  &  Full-disk   \\
\hline
\end{tabular}
\end{table*}

\section{Data analysis and results}
\begin{figure}
\centering
\includegraphics[width=0.9\linewidth,clip=]{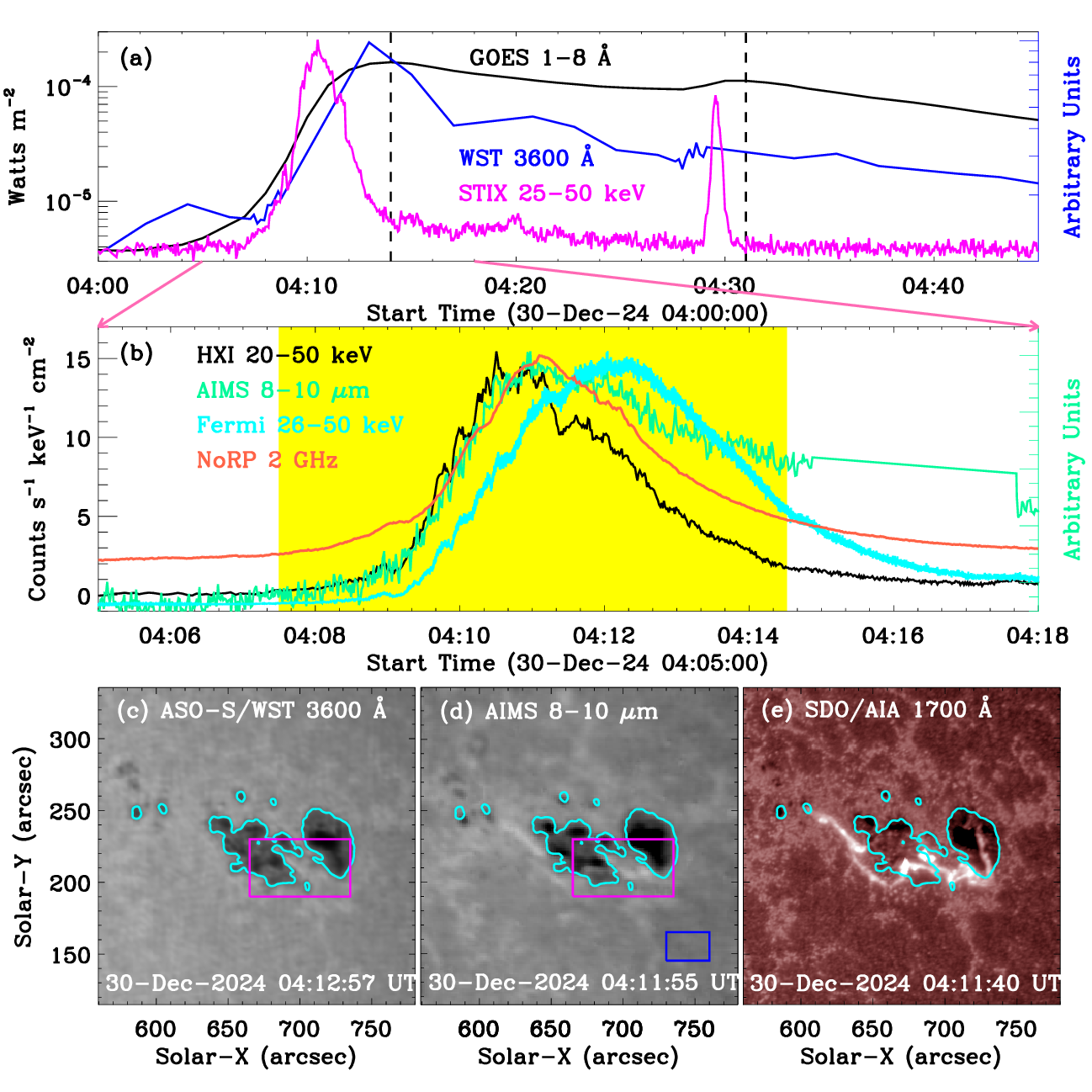}
\caption{Overview of the solar flare on 2024 December 30. (a): Light
curves during 04:00$-$04:45~UT measured by GOES~1$-$8~{\AA} (black),
WST~3600~{\AA} (blue), and STIX~25-50~keV (magenta). The vertical
lines mark the peak times of two solar flares in the GOES/XRS time
series. (b): Light curves from 04:05~UT to 04:18~UT measured in the
HXI~20$-$50~keV (black), AIMS~8-10~$\mu$m (spring green),
Fermi~26$-$50~keV (cyan), and NoRP~2~GHz (tomato-coloured)
wavebands. The yellow region marks the time frame on the Fourier
spectra. (c-e): Sub-maps with a FOV of about
220$^{\prime\prime}$$\times$220$^{\prime\prime}$ observed by WST,
AIMS, and AIA, respectively. The magenta rectangles outline the
local area used to integrate the white-light and mid-infrared
fluxes, while the blue rectangle marks the background region. The
cyan lines outline sunspots. \label{over}}
\end{figure}

Figure~\ref{over}~(a) presents the multi-wavelength time series
during the solar flare on 2024 December 30. The GOES~1$-$8~{\AA}
flux indicates two X-class flares (X1.5 and X1.1), peaking at about
04:14~UT and 04:31~UT, as marked by vertical lines. The
STIX~25$-$50~keV flux also reveals two peaks, confirming two
distinct flare events. The STIX flux was extracted from pixelated
science data, which had a time cadence of 3~s. Conversely,
white-light flux in the waveband of WST~3600~{\AA} shows one peak at
about 04:14~UT. With a routine 2-minute cadence, the WST flux was
integrated over the X1.5 flare region. While GOES and STIX fluxes
represent full-disk measurements, the WST flux corresponds to the
localized flare region (magenta rectangle). Thus, the two solar
flares originated from different active regions, with only the X1.5
event being studied. Figure~\ref{over}~(b) shows light curves in the
HXI~20$-$50~keV, Fermi~26$-$50~keV, NoRP~2~GHz, and AIMS~8-10~$\mu$m
wavebands during the X1.5 flare between 04:05$-$04:18~UT. The HXI
flux, derived from an open flux monitor, has a regular time cadence
of 1~s. Similarly, the AIMS flux, integrated over the flare region
(magenta rectangle) and normalized against the background time
series (blue rectangle), also has a regular time cadence of 1~s.
Conversely, the Fermi flux was interpolated into a uniform time
cadence of 1.024~s due to its irregularly sampled time series. They
all exhibit a number of successive and repeated wiggles superimposed
on a strong background trend, representing potential flare QPPs at
small amplitude. These small-amplitude QPPs are clearly evident in
AIMS, Fermi, and HXI data but not distinctly observed in NoRP
measurements. Figure~\ref{over}~(c)-(e) show local images measured
by ASO-S/WST, AIMS, and SDO/AIA, respectively. The sunspot regions
(outlined by cyan lines) enable cross-instrument alignment, because
all telescopes observed the same features.

To determine quasi-periods, we performed a fast Fourier transform
(FFT). This method was applied directly to the time series without
any further pre-processing, making it independent of detrending
procedures. The resulting power spectrum typically exhibits a
power-law distribution with a plateau, corresponding to red
noise and white noise, respectively. Thus, the power spectrum is
best modelled as a power-law function with a constant term ($C$)
\citep[cf.][]{Liang20,Inglis24}, as shown in Eq.~(\ref{eq1}):

\begin{equation}
\centering
 F (P) = A \cdot P^{\alpha} + C.
\label{eq1}
\end{equation}
\noindent Here, $F (P)$ represents the period distribution, $A$
denotes the amplitude, and $\alpha$ corresponds the power-law index.

\begin{figure}
\centering
\includegraphics[width=0.9\linewidth,clip=]{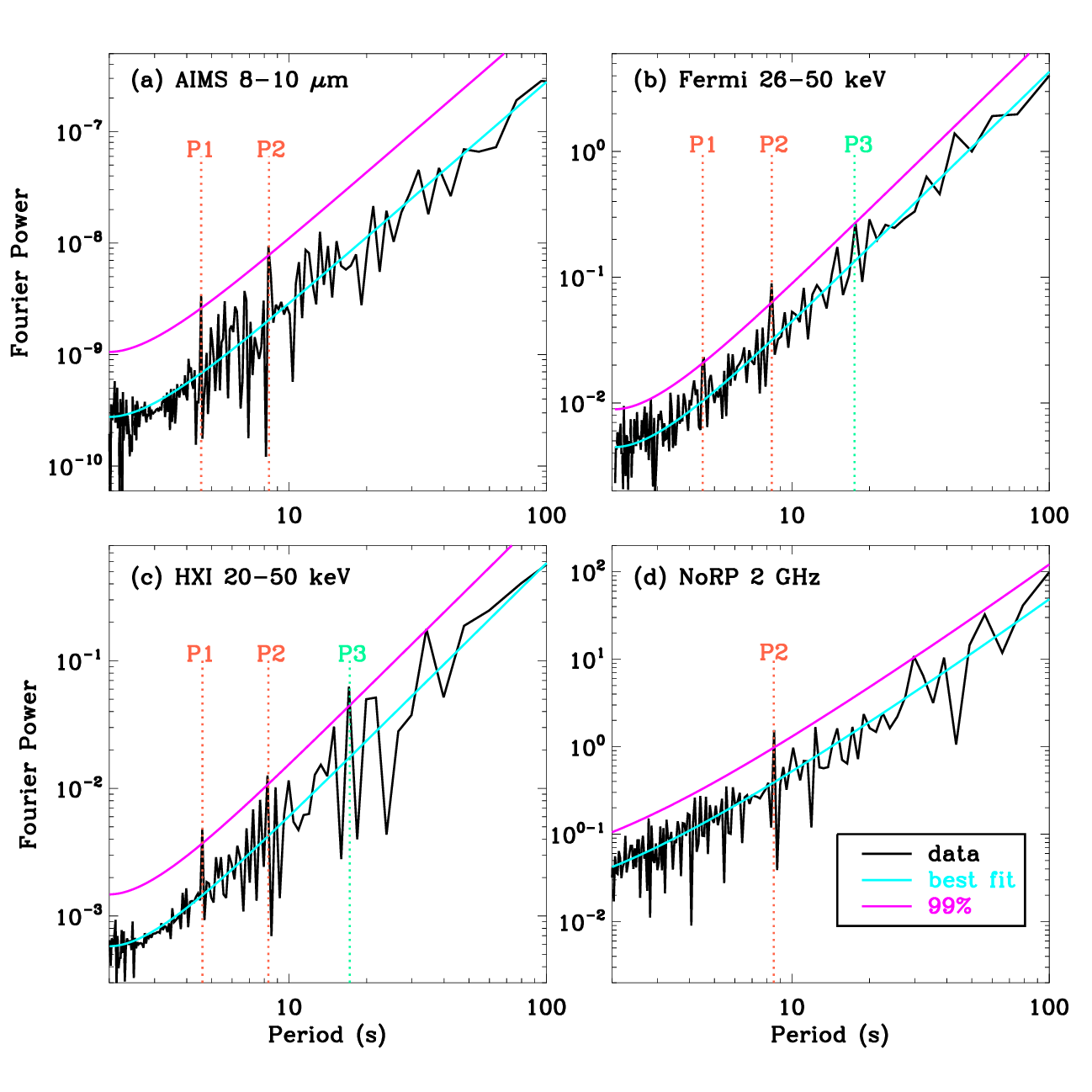}
\caption{Fourier power spectra in log-log space, obtained from the
AIMS~8-10~$\mu$m, Fermi~26$-$50~keV, HXI~20$-$50~keV, and NoRP~2~GHz
wavebands. The cyan and magenta lines in each panel indicate the
best fit and the 99\% confidence level, respectively. The dotted
lines outline interesting quasi-periods above the confidence level.
\label{fft1}}
\end{figure}

Figure~\ref{fft1} shows the Fourier power spectra derived from the
unfiltered time series in the AIMS~8-10~$\mu$m, Fermi~26$-$50~keV,
HXI~20$-$50~keV, and NoRP~2~GHz wavebands. Here, the time frame on
the Fourier spectra was from 04:07:30~UT to 04:14:30~UT, as
indicated by the yellow region in Fig.~\ref{over}~(b). The cyan and
magenta lines in each panel indicate the best fit and the 99\%
confidence level. We immediately note that two quasi-periods (P1 and
P2) exceed the 99\% confidence level in mid-infrared and HXR
emissions, corresponding to quasi-periods centred at about 8.5~s and
4.6~s, while only the longer quasi-period (P2) exceeds this level in
the microwave emission. Conversely, much longer quasi-periods ---
such as that centred at about 17.2~s (P3) --- are simultaneously
observed in HXR emission measured by HXI and Fermi but remain
undetected in mid-infrared emission. This might be due to the
different radiation mechanisms between HXRs and mid-infrared
emission. However, it is outside the scope of this study.

To investigate the double quasi-periods, we performed wavelet
analysis with a mother function of `Morlet' \citep{Torrence98} on the
detrended time series. The empirical mode decomposition (EMD)
technique \citep{Huang98} was employed to separate background trends
from detrended signals in the unfiltered time series, as this
approach requires no prior assumptions about trend characteristics.
The empirical background trend was identified as those modes with
periods exceeding specified fraction of the total signal duration
(denoted by a `cutoff' parameter), yielding a new set of modes with
periods shorter than this cutoff \citep{Kolotkov16,Li25a}. In our
case, the cutoff threshold was set to 0.2, meaning that a mode with
less than five oscillation cycles was reordered as the empirical trend.
This is because we wanted to strengthen the short-period components.
To compute the confidence level in the wavelet spectrum,
\cite{Torrence98} provided theoretical expressions assuming white-
and red-noise backgrounds. However, these assumptions may be
invalid, since the data had been detrended. The significance level
was therefore estimated by running a Monte Carlo (MC) simulation
\citep{Zhang12}.

\begin{figure}
\centering
\includegraphics[width=0.9\linewidth,clip=]{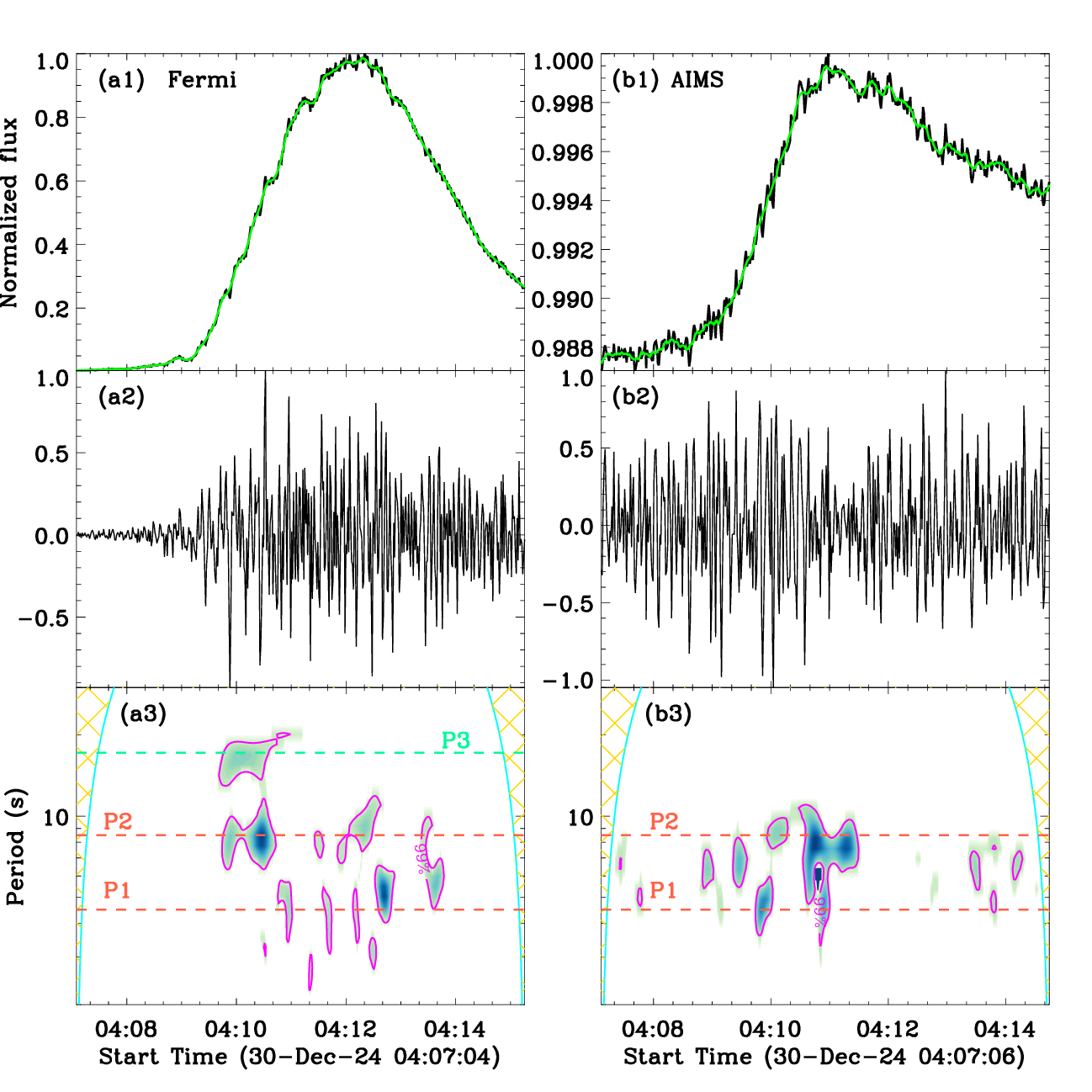}
\caption{Wavelet analysis results. (a1 and b1): Normalized light
curves and their trend measured by Fermi and AIMS. (a2 and b2):
Normalized detrended time series. (a3 and b3): Morlet wavelet power
spectra. The magenta contours outline the significance level at
99\%. The dashed lines mark the quasi-periods of interest.
\label{wav}}
\end{figure}

To emphasize the double-periodic QPPs, Figure~\ref{wav} presents the
`Morlet' wavelet analysis results in the HXR and mid-infrared
emissions measured by Fermi and AIMS. Panels~(a1) and (b1) show the
unfiltered time series, normalized by their maximum values. The
over-plotted green lines indicate their empirical background trends.
Panels~(a2) and (b2) show the detrended time series, normalized by
their maximum values. Some fluctuations are clearly evident in these
detrended series and may be regarded as QPP signals.
Figure~\ref{wav}~(a3) shows the wavelet power spectrum for the HXR
emission, which dominates by three bulks of power spectra within the
99\% significance level, centred at about 17.2~s, 8.5~s, and 4.6~s,
respectively. Figure~\ref{wav}~(b3) shows the wavelet power spectrum
for the mid-infrared emission, showing two bulks of power spectra
within the 99\% significance level, which are centred at about 8.5~s
and 4.6~s, respectively. This is consistent with the FFT power
spectra -- specifically, the centres labelled P1, P2, and P3.
Moreover, these periods mainly appear during the flare impulsive
phase.

\begin{figure}
\centering
\includegraphics[width=0.9\linewidth,clip=]{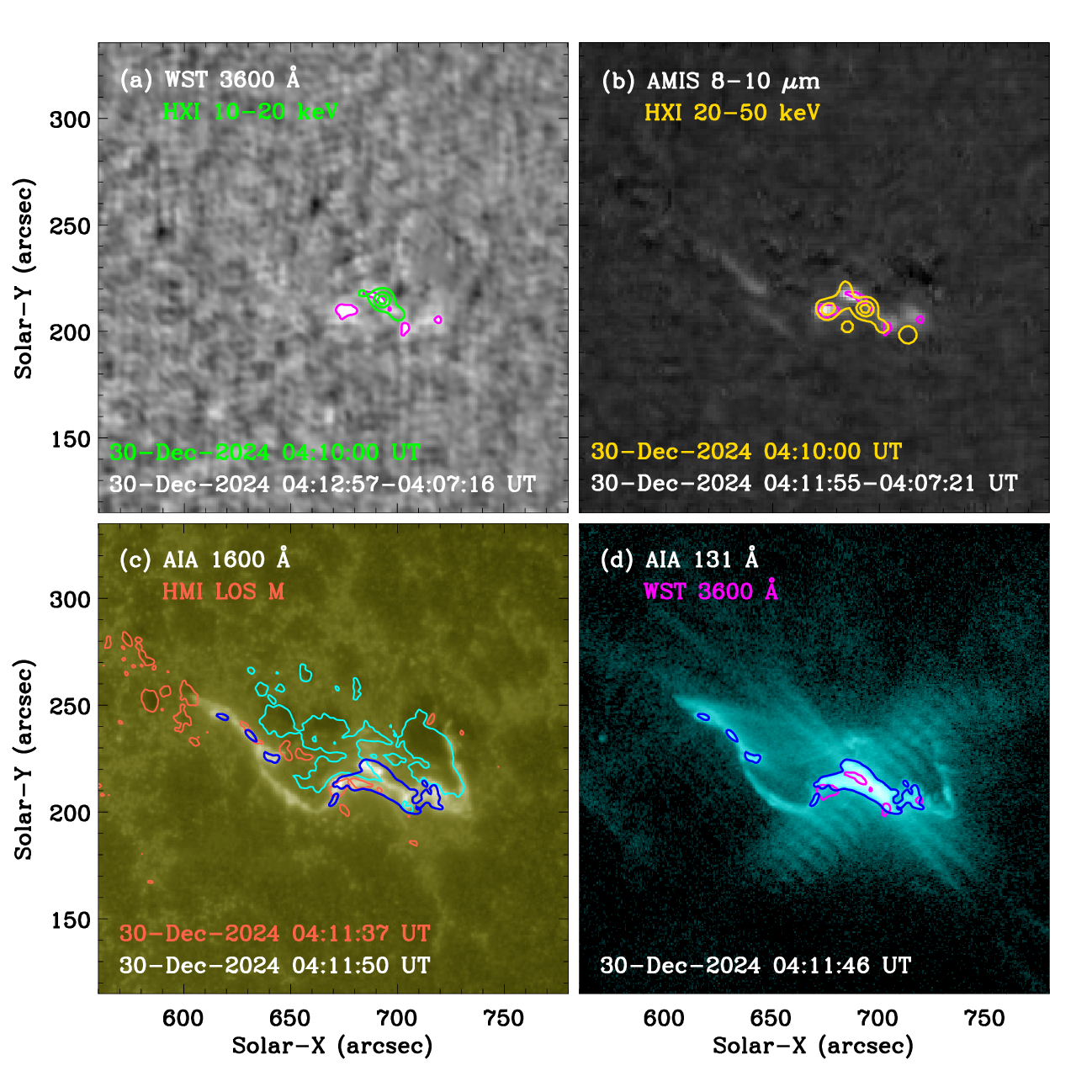}
\caption{Multi-wavelength images during the X1.5 flare. (a \& b):
Difference images in white-light and mid-infrared emissions measured by
WST and AIMS. The overlaid green and gold contours are reconstructed
from HXI data, and the levels are set at 20\%, 50\%, and 80\%,
respectively. (c \& d): UV/EUV sub-maps measured by AIA. The tomato-coloured
and cyan contours represent LOS magnetic fields at scales of
$\pm$500~G. The blue and magenta contours are from AIA~131~{\AA} and
WST~3600~{\AA}, respectively. \label{img}}
\end{figure}

To determine the source region of flare QPPs, we present
multi-wavelength images from the X1.5 flare in Fig.~\ref{img}.
Panels~(a) and (b) show difference images in the white-light and
mid-infrared wavebands, respectively. Difference imaging was used
instead of displaying the original images because white-light and
mid-infrared emissions during the X1.5 flare were relatively faint,
as shown in Fig.~\ref{over}~(c) and (d). The difference images
reveal four bright kernels in both white-light and mid-infrared
emissions. Three of these bright kernels are overlaid on HXR
emissions, which are outlined by gold contours. The HXR emission map
was reconstructed from the HXI data at 20-50~keV using the
HXI\_CLEAN algorithm, with a pixel scale of 2$^{\prime\prime}$. In
contrast, the SXR source obtained from HXI~10-20~keV consists of one
main source. Its location is consistent with one of the bright
kernels seen in the white-light and mid-infrared emissions, as
indicated by the green contour. Figure~\ref{img}~(c) and (d) show
the UV/EUV images at AIA~1600~{\AA} and 131~{\AA} wavelengths,
respectively. The tomato-coloured and cyan contours represent the
line-of-sight (LOS) magnetic field at levels of $\pm$500~G. The X1.5
flare exhibits several ribbon-like structures in the chromosphere
and transition region (i.e., AIA~1600~{\AA}), while revealing hot
loop-like features in the corona (i.e., AIA~131~{\AA}). The
multi-wavelength images suggest that the three bright kernels
observed in white-light and mid-infrared emissions are connected by
a high-temperate flaring loop, corresponding to the loop top and
double footpoints. Assuming a semicircular flaring loop profile
\citep{Tian16}, the loop length ($L$) can be estimated at about
21.5~Mm. The loop width ($w$), referred to as the full width at the
half maximum (FWHM) measured perpendicular to the loop axis, is
approximately 4.6~Mm.

\begin{figure}
\centering
\includegraphics[width=0.9\linewidth,clip=]{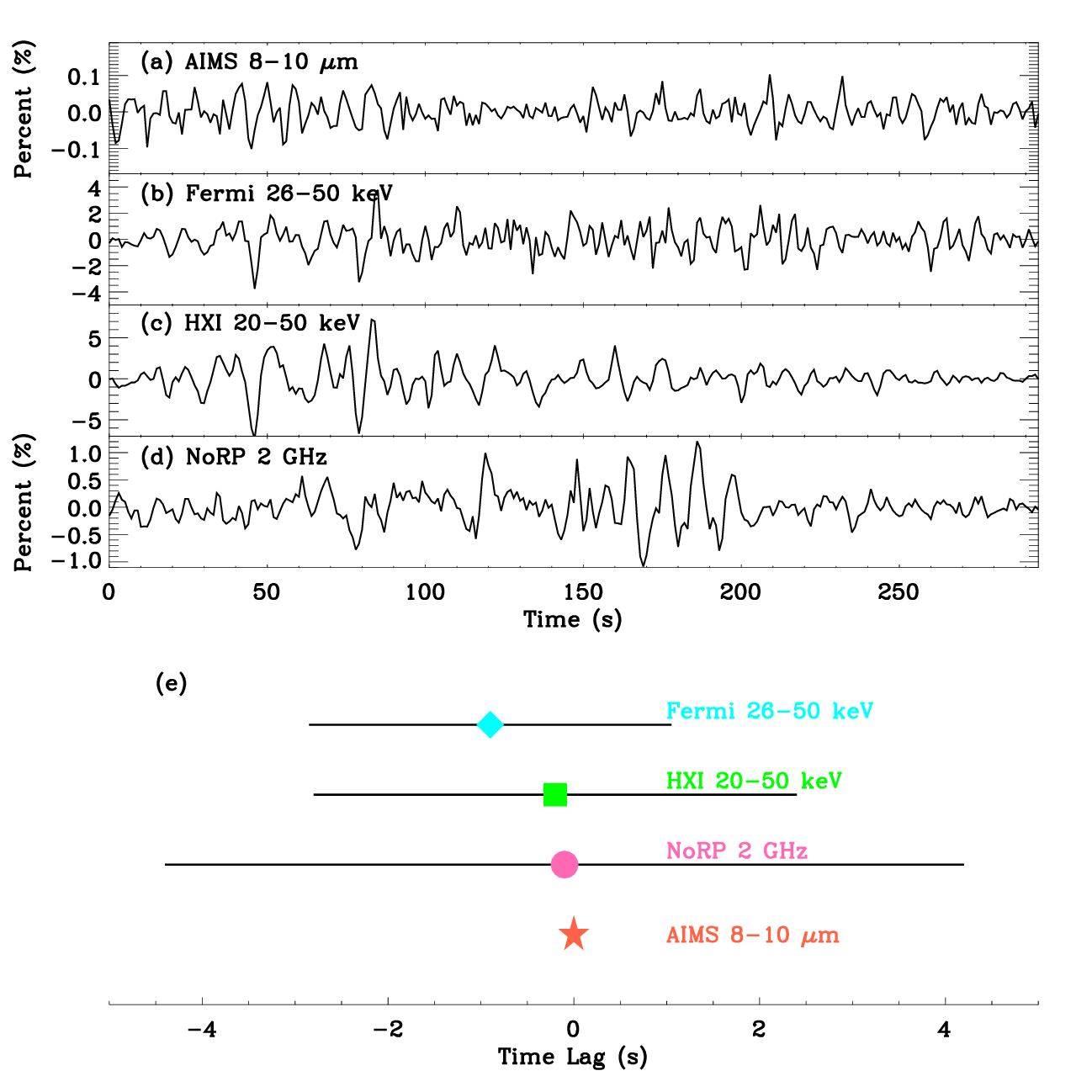}
\caption{Time delays for the detrended time series. (a)-(d):
Normalized detrended time series in multiple wavebands. (e): Time
delays with respect to that in AIMS~8-10~$\mu$m. The negative delay
means that the waveband leads the reference waveband. The coloured
symbol represents the centre of a Gaussian fit to the cross-correlation
function, and the horizontal lines represent the FWHM of the
Gaussian fit. \label{delay}}
\end{figure}

The correlation of pulse peak timings in multiple wavebands during a
solar flare contains information about the flaring energy release
and deposition processes. To compare the time lag of flare QPPs in
various wavebands, we performed cross-correlation analysis on the
QPP patterns detected in different wavebands, as shown in
Fig.~\ref{delay}. Panels~(a)-(d) show the time series of the
modulation depth in the mid-infrared, HXR, and microwave emissions
wavebands. Here, the modulation depth ($d_M$) is defined as the
ratio between the detrended time series ($dt$) and their background
trends ($bt$) following Eq.~(\ref{eq2}):

\begin{equation}
\centering
 d_M = \frac{dt}{bt}~\ast~100\%.
\label{eq2}
\end{equation}

It is notable that pulse peak timings of mid-infrared emission
(i.e., AIMS~8-10~$\mu$m) are closely aligned with those of
nonthermal emissions seen in the Fermi~26-50~keV, HXI~20-50~keV, and
NoRP~2~GHz wavebands. Also, the modulation depth of infrared QPPs is
less than 0.1\%, while those in HXR and microwave emissions are in
the range of 1\%-5\%, indicating low-amplitude QPP signals.

Figure~\ref{delay}~(e) presents the time delay relative to the
AIMS~8-10~$\mu$m waveband, where a negative delay indicates that the
waveband leads the reference. The time delay is defined as the
centre of a single Gaussian fit to the cross-correlation function,
with the duration represented by the FWHM of that Gaussian function,
as indicated by the coloured symbols and horizontal lines.
Cross-correlation analysis results show that the delay time is less
than 1~s, which is smaller than the time resolution of the
observational instruments. Thus, the time delay can be negligible,
supporting the conclusion that the QPP pattern seen in the
mid-infrared emission is strongly associated with those seen in HXR
and microwave emissions. The broader duration of the
cross-correlation function in microwave emission (nearly twice that
in HXR and mid-infrared emissions) occurs primarily since only the
longer quasi-period (P2) is detected in NoRP~2~GHz. The close
association implies that mid-infrared brightening during the X1.5
flare may depend on bombardment by nonthermal electrons accelerated
through magnetic reconnection.

\begin{figure}
\centering
\includegraphics[width=0.9\linewidth,clip=]{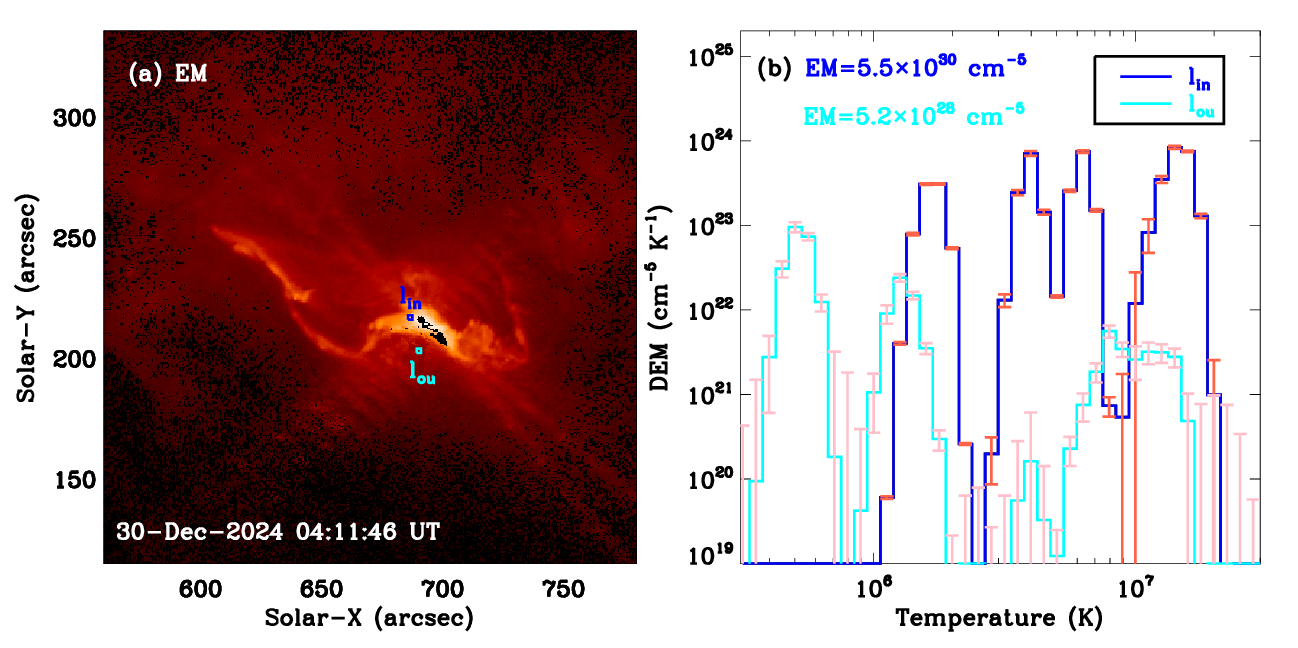}
\caption{DEM analysis results. (a): EM map integrated in the
temperature range of 0.31-30~MK. The blue and cyan boxes mark the
regions inside ($l_{in}$) and outside $l_{ou}$ the flaring loop.
(b): DEM profiles and their uncertainties at the selected regions.
\label{dem}}
\end{figure}

Figure~\ref{dem} presents differential emission measure (DEM)
analysis results for the X1.5 flare, derived from AIA images in six
EUV wavelengths. Using the improved sparse-inversion code
\citep{Cheung15,Su18}, we calculated the distribution of DEM(T) for
every pixel, estimating uncertainties from 100 MC simulations as
three times the standard deviation of these realizations.
Figure~\ref{dem}~(a) shows the emission measure (EM) map integrated
over 0.31-30~MK. Similar to the AIA~131~{\AA} map
(Fig.~\ref{img}~d), the EM map reveals a high-temperature plasma
loop, with black patches corresponding to saturated pixels in AIA
EUV images. We selected two small regions with a field of view (FOV)
of about 1.8$^{\prime\prime}$$\times$1.8$^{\prime\prime}$ to display
DEM profiles inside ($l_{in}$) and outside ($l_{ou}$) the flaring
loop. Panel~(b) depicts the DEM profiles as a function of plasma
temperature (DEM(T)), and the error bars (colour-coded) represent
the MC-derived uncertainties. The EM values were calculated inside
and outside the flaring loop, as labelled accordingly. Considering a
LOS integration length equal to the loop width ($w$) and a filling
factor of 1.0 \citep{Tian16}, we calculated the number density
($n_{ei}$) inside the flaring loop using Eq.~\ref{eq3}:

\begin{equation}
\centering
 n_e = \sqrt{EM/w},
\label{eq3}
\end{equation}
\noindent where, $n_e$ is the number density, $w$ represents the LOS
integration length along the flaring loop. Our calculation yields
$\sim$1.1$\times$10$^{11}$~cm$^{-3}$.

For the background corona outside the loop (devoid of loop
structures), we adopted an effective LOS depth \citep[cf.
$w~\approx~4~\times~10^{10}~cm$,~][]{Zucca14} to estimate the number
density ($n_{eo}$) outside the flaring loop, obtaining
$\sim$1.1$\times$10$^{9}$~cm$^{-3}$. This results in a density
contrast of roughly 100, indicating that the flaring loop is
substantially denser than its surroundings.

\section{Conclusions and discussions}
We analysed an X1.5 flare on 2024 December 30, simultaneously
observed by AIMS, WST, Fermi, HXI, STIX, NoRP, AIA, and HMI across
the mid-infrared, white-light, HXR, microwave, and UV/EUV wavebands. The
solar flare exhibits significant enhancement in the
WST~3600~{\AA} waveband, indicating a white-light flare \citep{Liy24,Song25}.
Imaging observations from AIMS~8-10~$\mu$m in the high-cadence mode
provide an unprecedented opportunity to study the infrared
behaviour of QPPs \citep{Deng25}. Combined with HXR and microwave data, we
can explore the origin of infrared QPPs. Conversely, the white-light
data from WST in the routine mode exhibit insufficient cadence, making
it impossible to compare the short period in white-light and
mid-infrared emissions. Using the FFT and wavelet analysis method,
we first identified dual quasi-periods centred at about 8.5~s and
4.6~s in AIMS~8-10~$\mu$m data. These were also detected in
Fermi~26-50~keV and HXI~20-50~keV data, while only the longer
quasi-period appeared in NoRP~2~GHz data. This multi-instrument
detection of dual periods rules out instrumental artefacts. The
modulation depths of flare QPPs are remarkably small -- 1\%-5\% in
the HXR and microwave emissions and $<$ 0.1\% in the mid-infrared emission --
indicating low-amplitude oscillatory signals.

Flare QPPs were observed across nearly all wavebands, including
radio, UV/EUV, white light, SXR/HXR, and even $\gamma$-rays
\citep[e.g.,][]{Nakariakov10,Nakariakov18,Tan10,Tan16,Li24a,Li24b,Ashfield25,French25,Lim25,Song25,Cattell26}.
However, flare QPPs in infrared emission remained unreported prior
to this study due to observational limitations. These limitations
included the insufficient sensitivity of existing solar infrared
telescopes, strong atmospheric interference affecting ground-based
observations, and the limited number of space-based infrared
observatories. Here, we report the first detection of
double-periodic QPPs in mid-infrared emission during the impulsive
phase of an X1.5 flare. The modulation depth, which is determined by
the detrended time series and its empirical background trend, is
less than 0.1\%. Similar minimal modulation depth has been detected
in white-light QPPs \citep{Li24b}, indicative of weak oscillatory
signals. This phenomenon arises because solar background radiation
dominates in infrared and white-light wavebands, while flare-induced
enhancements remain comparatively faint
\citep{Miteva16,Li24b,Yang25}, resulting in the
reduced modulation depth. This also explains the scarcity of
reported infrared QPPs. Furthermore, the quasi-period of white-light
QPPs is often quite long, i.e., $\sim$8~minutes
\citep{Zhao21,Li24b,Li25b}, constrained by instrumental time-cadence
limitations. While flare QPPs with quasi-periods less than 10~s are
frequently detected in HXR and microwave emissions, owing to higher
time cadences and instrument sensitivity \citep{Inglis24,Li25}.
In this study, the flare QPPs at short quasi-periods (i..e, $<$10~s)
are simultaneously detected the mid-infrared, HXR and
microwave wavebands. Crucially, the co-detection of dual quasi-periods in HXR
and mid-infrared emissions suggests a nonthermal origin of flare
QPPs. In contrast to the predominantly thermal excitation of
mid-infrared emission during solar flares
\citep{Simoes17,Lopez22,Yang25}, our results strongly suggest a
primarily nonthermal origin for the detected infrared QPPs
\citep[e.g.,][]{Penn16} .

It is necessary to discuss the generation mechanism of the infrared
QPPs exhibiting double quasi-periods. Imaging observations show that
mid-infrared emission is spatially co-located with white-light and
HXR emissions. Based on the standard 2D flare model
\citep{Priest02}, flare radiation in white-light and HXR wavebands
is closely associated with nonthermal electrons accelerated by
magnetic reconnection \citep{Yang25b}. In view of the above, we
deduce that the flare QPPs observed in mid-infrared emission are
likely triggered by a quasi-periodic regime of oscillatory magnetic
reconnection \citep[e.g.,][]{Kumar24,Zhu25,Zimovets25}. Now, the
question is whether the periodic reconnection is spontaneous or
induced \citep{Ashfield25,Li26}. In the standing-mode MHD wave, the
phase speed ($C_{ph}$) is determined by the loop length ($L$), the
quasi-period ($P$), and the harmonic number (j) of the standing
mode, as shown in Eq.~\ref{eq4} \citep{Li20,Wang21}:

\begin{equation}
\centering
 C_{ph} = \frac{2L}{jP}.
\label{eq4}
\end{equation}

If we consider the longer quasi-period (P2) at about 8.5~s detected
in mid-infrared emission, the phase speed for the fundamental number
is estimated to about 5000~km~s$^{-1}$. Even when considering a much
longer quasi-period (P3) of about 17.2~s observed in HXRs, the phase
speed remains as high as about 2500~km~s$^{-1}$. The phase speed is
much higher than the local sound speed of $\sim$500~km~s$^{-1}$ in
hot ($\sim$11~MK) flaring loops \citep{Wang21}, and it also
significantly exceeds the average kink speed of about
1328~km~s$^{-1}$ \citep{Nakariakov21}. Consequently, the flare QPPs
cannot be modulated by standing MHD waves in either slow or kink
modes. Rather, the estimated phase speed aligns with the range for
fast sausage-mode waves, i.e., $\sim$2400-5000~km~s$^{-1}$
\citep{Tian16,Li20}. It is well known that sausage-mode waves
readily propagate in broader, denser plasma loops.
\cite{Nakariakov03} derived a necessary condition for sausage-mode
waves in solar coronal environments, as shown in Eq~\ref{eq5}:

\begin{equation}
\centering
 \frac{n_{ei}}{n_{eo}}~>~(\frac{L}{0.65w})^2,
\label{eq5}
\end{equation}
\noindent where, $n_{ei}$ and $n_{eo}$ denote the number densities
inside and outside the flaring loop, respectively, while $L$ and $w$
represent the loop length and the loop width, respectively.

In our case, the necessary condition threshold is about 52, consistent
with the value reported by \citet{Tian16}. Based on our DEM results and
morphological analysis, the density contrast between the interior
and exterior of the flaring loop is estimated to about 100. This
significantly exceeds the required threshold, implying that
sausage-mode waves can propagate within the flaring loop. The double
quasi-periods at about 8.5~s and 4.6~s are probably modulated by
quasi-harmonics of sausage-mode waves in this flaring loop. However,
it remains challenging to determine the exact number of
quasi-harmonics, due to the multiple quasi-periods
detected in HXRs. Sausage oscillations in hot plasma loops have
previously explained flare QPPs observed in the
radio, SXR/HXR, Ly$\alpha$, and mid-ultraviolet Balmer continuum wavebands
\citep[e.g.,][]{Van11,Kolotkov18,Li21,Ning22}, despite observational
difficulties in solar coronal conditions where the plasma loop must
be sufficiently thick and dense \citep{Nakariakov03}. Based on
spectroscopic observations with high spatio-temporal resolution,
global sausage waves have been identified in hot plasma loops
\citep{Tian16,Ashfield25}. These findings support the presence of
sausage waves in the solar corona. Therefore, the double-periodic
pulsations may be triggered by oscillatory magnetic reconnection
modulated by quasi-harmonics of sausage-mode waves.

\section{Summary}
Based on a suite of state-of-the-art ground-based and space-borne
observatories, including AIMS, Fermi, HXI, STIX, NoRP, WST, AIA, and
HMI, we investigated double-periodic QPPs during the X1.5 flare on
2024 December 30. Our key findings are summarized below:

(1) High-cadence AIMS observations enabled the first detection of
dual quasi-periods centred at about 8.5~s and 4.6~s in mid-infrared
emission. Additionally, double-periodic pulsations were observed in
HXRs by both Fermi and HXI, while only the longer period appeared in
microwave emission.

(2) The mid-infrared sources were spatially matched with source
regions in white-light and HXR wavebands during the flare impulsive
phase, indicating a close association between mid-infrared radiation
and nonthermal electrons produced by magnetic reconnection.

(3) The flare QPPs across mid-infrared, HXR, and microwave wavebands
may be associated with nonthermal electrons that are periodically
accelerated by oscillatory magnetic reconnection, with double
periods modulated by quasi-harmonics of fast sausage waves.

\begin{acknowledgements}
The authors would like to thank the referee for inspiring comments.
This work is funded by the National Key R\&D Program of China
2021YFA1600502 (2021YFA1600500), NSFC under No.12573057, and the
Strategic Priority Research Program of the Chinese Academy of
Sciences, Grant No. XDB0560000. X.~Yang is supported by Xinjiang
Leading Talent Introduction Project, Grant No.
XJRC-2025-KJ-YJ-CXPT-066. We thank the teams of AIMS, ASO-S, Fermi,
STIX, NoRP, SDO, and GOES for their open data use policy.
\end{acknowledgements}

\end{document}